\documentclass[referee]{aa} % for a referee version
\usepackage{graphicx}
\def\AJ{AJ }
\def\ApJ{ApJ }
\def\AA{A\&A }
\def\AAS{A\&AS }
\def\ARAA{ARA\&A }

\def\ApJS{ApJS }

\def\MNRAS{MNRAS }
\def\degree {\ifmmode^\circ\else$^\circ$\fi}                       % degree
\def\kms{\ifmmode {\,{\rm km\,s^{-1}}}                          % km s-1
        \else {\hbox{$\,$ {\rm km$\,$s$^{\rm -1}$}}}\fi}
\begin{document}
   \thesaurus{06     % A&A Section 6: Form. struct. and evolut. of stars
              (08.06.2
              09.09.1
              09.10.1
              09.13.2
              13.18.8)}
   \title{The origin of the HH\,7--11 outflow\thanks{Based on
   observations carried out with the IRAM Plateau de Bure
   Interferometer. IRAM is supported by INSU/CNRS (France), MPG
   (Germany) and IGN (Spain).}}

%  \subtitle{I. Overviewing the $\kappa$-mechanism}

   \author{R. Bachiller\inst{1}, F. Gueth\inst{2}, 
           S. Guilloteau\inst{3}, 
           M. Tafalla\inst{1}, \and A. Dutrey\inst{3}}
%                              \and
%                      C. Ptolemy\inst{2}\fnmsep\thanks{Just to show the usage
%                      of the elements in the author field}

   \offprints{R. Bachiller}

   \institute{IGN Observatorio Astron\'omico Nacional, 
              Apartado 1143, E-28800 Alcal\'a de Henares, Spain\\
              email: bachiller@oan.es
         \and
            Max-Planck-Institut f\"ur Radioastronomie, 
            Auf dem H\"ugel 69, D-53121 Bonn, Germany        
         \and
             Institut de Radio Astronomie Millim\'etrique, 
             300 rue de la Piscine, F-38406 Saint Martin d'H\`eres, France}
%
%             email: c.ptolemy@hipparch.uheaven.space
%             \thanks{The university of heaven temporarily does not
%                     accept e-mails}

\authorrunning{Bachiller et al.}
%\titlerunning{HH\,7--11}

   \date{Received: August 2000; accepted: September 2000}

   \maketitle

\begin{abstract}

New, high-sensitivity interferometric CO $J$=2--1
observations of the HH\,7--11 outflow show that despite previous
doubts, this system is powered by the Class~I source SVS\,13. The
molecular outflow from SVS\,13 is formed by a shell with a large
opening angle at the base, which is typical of outflows from Class~I
sources, but it also contains an extremely-high-velocity jet composed
of ``molecular bullets'', which is more typical of Class~0
outflows. This suggests that SVS\,13 could be a very young Class~I,
which still keeps some features of the previous evolutionary stage. We
briefly discuss the nature of some sources in the SVS\,13 vicinity
which are emitters of cm-wave continuum, but have no counterpart at mm
wavelengths.

%Radio lines: molecular: interstellar}

   \end{abstract}
%________________________________________________________________

%----------------------------------------------------------------------------

\begin{figure*}[t]
\centerline{\includegraphics[angle=-90,width=16cm]{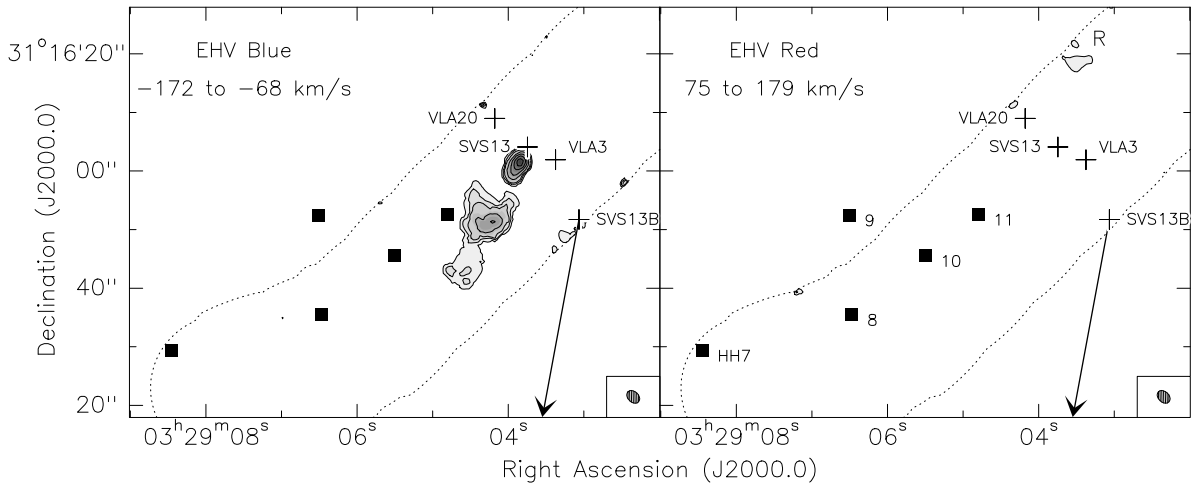}}
\caption{CO $J$=2--1 line intensity around HH\,7--11, integrated in 
extremely-high-velocity (EHV) intervals of 104{\kms} width.
The LSR velocity intervals are given in the top left corners.
Contours are 1.4, 2.8, 4.2 Jy\,beam$^{-1}${\kms}, and then increase
by 2.8 Jy\,beam$^{-1}${\kms} step.  Crosses mark the positions of VLA sources
(Rodr\'{\i}guez et al. 1997, 1999), and of SVS\,13 and SVS\,13B
(Bachiller et al. 1998).  The arrow indicates the direction of the
SVS\,13B molecular outflow, and filled squares mark the positions of
HH objects. The dotted line shows the limits of the observed region.
The clean beam is also indicated.
}
\label{}
\end{figure*}

%----------------------------------------------------------------------------

\section{Introduction}

HH\,7--11 is one of the most conspicuous chains of Herbig-Haro (HH)
objects. It lies $\sim 6\arcmin$ South from NGC1333, in a region
crowded of young stellar objects (YSOs) and low-mass star-formation
signspots (bright and dark nebulosities, jets, OH and water masers,
etc.). The region is at only $\sim$300 pc from the Sun (Herbig \&
Jones 1983, Cernis 1992).  Overall views in the optical and the
infrared have been reported by Aspin et al.\ (1994), Hodapp \& Ladd
(1995) and Bally et al.\ (1996).

HH\,7--11 is the optically visible part of an energetic high-velocity
bipolar outflow which has bright emission in CO (Snell \& Edwards
1981, Bachiller \& Cernicharo 1990, Masson et al.\ 1990, Knee \&
Sandell 2000), H$_2$ (Hoddap \& Ladd 1995),
HI (Lizano et al. 1988, Rodr\'{\i}guez et al. 1990), and SiO (Lefloch
et al. 1998, Codella et al. 1999).  In spite of so many detailed
studies, the HH\,7--11 area is so rich in YSOs that identifying the
outflow driving source is not an easy task. For as long as 20 years
the infrared star SVS\,13 (Strom et al. 1976), which is approximately
aligned with the HH\,7--11 string, was believed to be its exciting
source. However, recent radio observations (Rodr\'{\i}guez et
al. 1997, 1999) have revealed new radio sources close to SVS\,13 and
have casted doubts about which is the source actually driving the
outflow. Rodr\'{\i}guez et al. (1997) favored VLA\,3, a source placed
6{\arcsec} SW from SVS\,13 and also well aligned with the HH\,7--11
chain, as the most likely exciting source of the outflow. The region
is made even more complex by the presence of two other nearby sources,
the radio source VLA\,20 (Rodr\'{\i}guez et al.\ 1999) and the Class~0
protostar SVS\,13B (Bachiller et al.\ 1998). The latter is known to
drive a highly collimated SiO jet (Bachiller et al.\ 1998). In this
Letter we report interferometric CO $J$=2--1 observations showing
unambiguously that the actual driving source of the HH\,7--11 flow is
SVS\,13. Our observations also provide important information on the
structure of this prototypical outflow.

%-----------------------------------------------------------------------------

\section{Observations}

The observations were carried out in March 1998 with the IRAM
5-antenna interferometer at Plateau de Bure. Three configurations of
the array were used, with baselines extending up to 176~m.  The
dual-channel receivers were tuned to the CO $J$=2--1 and SiO $J$=2--1
frequencies.  The SiO and continuum observations have been presented
by Bachiller et al.\ (1998).  Typical SSB system temperature was 300 K
at the CO frequency. The correlator was configured to give a
resolution of 3.3\kms\ in a $\sim$400\kms\ wide interval.  Phase
calibration was achieved by observing 3C84, which is close in the sky
to HH\,7--11. Typical rms phase noise on the longest baselines was
30$\degree$. To correct for decorrelation due to the phase noise, the
amplitude was also calibrated through the observations of 3C84, whose
flux density was 2.6~Jy at 230.5~GHz. A mosaic of 10 fields was
observed in order to cover the central region of the HH\,7--11
outflow. Interferometric images were produced using natural weighting
and cleaned with the GILDAS software. The clean beam is
2.4$\times$1.8{\arcsec} at the CO frequency. The continuum emission
(see Bachiller et al. 1998) has been subtracted from the line maps.

%-----------------------------------------------------------------------------

\section{Results}

The interferometric CO $J$=2--1 images of the HH\,7--11 vicinity are
complex, reflecting the source crowding in the area. In order to
summarize our results, we have divided the high-velocity emission into
a limited number of velocity intervals. The most extreme velocities
are presented in Fig.~1.  Following Bachiller \& Cernicharo (1990) we
will refer to this component as the ``extremely high velocity'' (EHV)
component, whereas the ``standard high velocity'' (SHV) component will
designate the emission at lower velocities shown in Fig.~2. There is
some emission at intermediate velocities exhibiting an intermediate
behavior as compared to the maps discussed here.

%----------------------------------------------------------------------------

\begin{figure*}[t]
\centerline{\includegraphics[angle=-90,width=16cm]{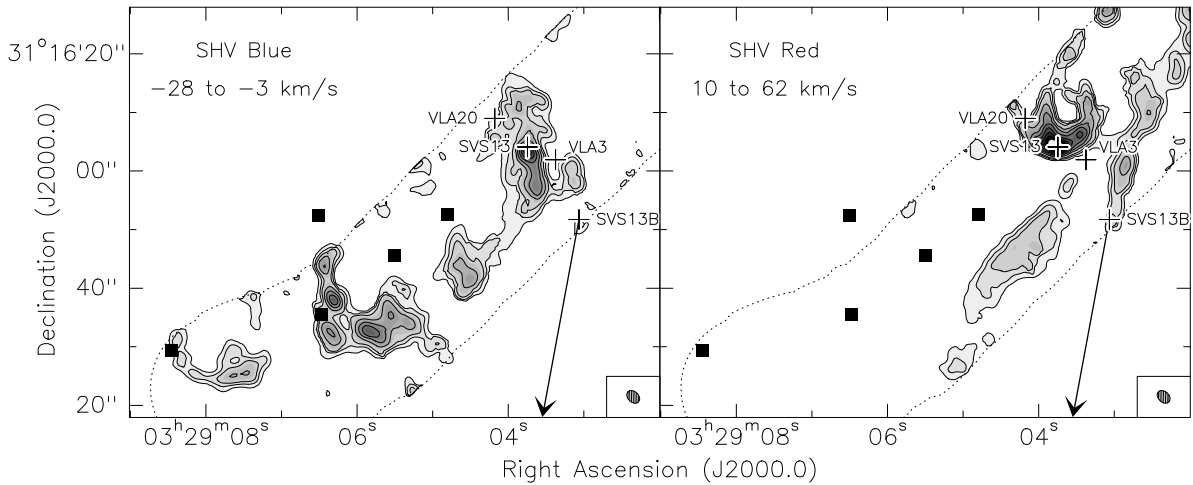}}
\caption{CO $J$=2--1 intensity integrated in the
standard-high-velocity (SHV) intervals which are indicated in the top
left corner (velocities with respect to the LSR).  Contours and
symbols are as in Fig.~1.}
\label{}
\end{figure*}

%----------------------------------------------------------------------------

\subsection{EHV emission: the origin of HH\,7--11}

The EHV emission (Fig.~1) reveals an extended blueshifted structure,
$\sim$25{\arcsec} long in the southeast direction from SVS\,13, in
good agreement with previous lower resolution maps (Bachiller \&
Cernicharo 1990, Masson et al. 1990). This jet-like feature consists
of three well aligned clumps (``molecular bullets'') and emanates
exactly from the position of SVS\,13.  As can be seen in the right
panel of Fig.~1, the EHV redshifted emission reveals a small clump
(marked ``R'' in the figure), $\sim$15{\arcsec} to the northwest
from SVS\,13. This bullet is well aligned with the blueshifted jet
and is therefore likely to trace the redshifted, weaker lobe. The
lower resolution maps already showed that the EHV redshifted emission is
relatively weak in this area. There is however a strong peak
$\sim$45{\arcsec} northwest from SVS\,13, well beyond the area we
imaged.

Bullet R is well detected over the noise level: its CO spectrum is a
feature of $\sim$10{\kms} linewidth, centered at V$_{\mathrm
LSR}$ =147{\kms}.  The LSR velocities of the three blueshifted CO
bullets are $-$70, $-$110, and $-$150{\kms}, when moving from SVS\,13
to the southeast. An analysis of the extreme velocity SiO $J$=2--1
emission (not shown in Bachiller et al. 1998) shows that the first CO
bullet also presents weak SiO emission in the same velocity range
(i.e. up to $-$80{\kms}). The SiO emission is compact ($< 4${\arcsec})
and has an integrated intensity of 1.8 Jy{\kms}. The CO and SiO
emission peaks are coincident within 0.5{\arcsec}, and the CO/SiO
integrated intensity ratio is $\sim$25.  Unfortunately, because of
backend limitations, the velocity coverage was insufficient to observe
in SiO the two CO bullets more distant from SVS\,13.

The typical sizes of the blueshifted CO bullets are $\sim$5{\arcsec},
or 1500~AU at the assumed distance of 300~pc.  The distance between
successive CO peaks is $\sim$10{\arcsec} ($\sim$3000 AU). By assuming
that the outflow is inclined by about 45{\degree} to the plane of the
sky, we find that the time elapsed between successive ejection events
is $\sim$100 yr. In this context, it is worth noting that SVS\,13
itself also displays a high variability (Eisloffel et al. 1991), so it
is tempting to suggest that the eruptions of the outflow are related
to outbursts in the driving source.

The high spatial coincidence of the EHV jet extremity with SVS\,13
(seen in both CO and SiO emission), and the
lack of significant high-velocity CO emission from VLA3 and VLA20,
strongly indicate that SVS\,13 is the actual driving source of the
HH\,7--11 outflow.

\subsection{SHV shell-like outflow}

The SHV maps provide interesting information on the SVS\,13B and
SVS\,13 outflows.  The red map uncovers the northern lobe from
SVS\,13B, which is observed to be remarkably well aligned with the
southern blueshifted one (marked with an arrow in the figures; see
Bachiller et al. 1998) and exhibits a similar high collimation. A SiO
clump in this redshifted jet was reported by Bachiller et al. (1998),
and coincides with the closest CO peak from SVS\,13B.

Around HH\,7--11, the SHV blueshifted emission forms a well developed
arc, strongly suggesting it traces the southern flank of a shell
surrounding the HH objects, with the apex at SVS\,13. This SHV shell
is coincident with the nebula seen in the optical and near-IR images
(e.g. Hodapp \& Ladd 1995), corresponding well to the cavity where the
HH objects are formed. This is illustrated in Fig. 3 where the SHV
emission is overlaped on the K$'$ image of Hodapp (1994). The shell
structure is not smooth, but appears broken in irregular clumps, which
could be created as a result of the interaction with a strongly
inhomogeneous medium, or by some instability process. Interestingly,
there is a bow-shaped SHV structure near HH\,7, probably indicating
that this object marks an extremity of the cavity.

%----------------------------------------------------------------------------

\subsection{Changes in the direction of the outflow axis}

The blueshifted EHV jet lies at P.A. 155{\degree}, i.e. on a different
direction from that defined by the HH objets and the SHV shell (P.A.
$\sim$130{\degree}). Since the EHV jet has probably been created by
the most recent ejection phenomena (as indicated by its shorter
kinematical timescale), we conclude that the outflow axis has moved by
about $\sim$25{\degree} toward the South from the period in which the
SHV shell was created to the most recent ejection events. This change
of the outflow axis could be due to precession in the ejection or to
deflection of the outflow along its path.  Recent observations of the
H$_2$ shocks show that the proper motion vectors near HH\,11 and
HH\,10 (the closest HH objects to SVS\,13) have higher P.A. than those
near HH\,7 and HH\,8 (Chrysostomou et al. 2000, their Fig.~1). The
proper motions of HH\,10--11 are very well aligned with the CO EHV jet
and point to SVS\,13.  However, the proper motion vectors of HH\,7--8
are better aligned with the SHV shell, and point to a position
$\sim$20{\arcsec} south from SVS\,13.  Since a precession effect
should preserve the vectors pointing towards SVS\,13, the observed
changes in the flow direction are probably due to the action of a
force lateral to the outflow (e.g. dynamical or magnetic pressure, or
collisions with ambient dense clumps).
%----------------------------------------------------------------------------

\section{Discussion}

\subsection{Evolutionary status of the HH\,7--11/SVS\,13 system}

The data presented here provide important information on the structure
of the HH\,7--11 molecular outflow, and direct evidence on the
identity of its driving source (SVS\,13). It thus appears interesting
to connect the characteristics of this well studied YSO with the
properties of its outflow.

SVS\,13 is a strong infrared source (Strom et al. 1976) of
85~L$_\odot$ (based on Molinari et al. 1993, scaled to the distance
assumed here) which experienced outbursts between 1988 and 1990
(Eisloffel et al. 1991) becoming even optically visible (Mauron \&
Thouvenot 1991). The source is associated with a cm-wave source (named
VLA4, Rodr\'{\i}guez et al. 1999) and it is embedded in a dusty disk
or elongated envelope, of $\sim$1~M$_\odot$, which is roughly
perpendicular to the HH\,7--11 chain (Bachiller et al. 1998). Briefly
speaking, the characteristics of SVS\,13 are those of a Class~I object
undergoing eruptions.

The HH\,7--11 shell structure is similar to shells observed in both
outflows from Class I sources (e.g. L\,1551/IRS\,5, Moriarty-Schieven et al. 1987)
and outflows from Class 0 sources  (e.g. L1527 IRS: Tamura et
al. 1996; L1448: Bachiller et al. 1995).  
The L\,1551 outflow is, however, spatially much more extended than HH\,7--11,
indicating that L\,1551 is more evolved. Which is most notable in
the case of SVS\,13 -in principle a Class\,I object- is the presence
of an EHV jet. Up to now such jets
were believed to be characteristics of Class~0 sources such as
L\,1448-mm or IRAS\,03282+3035 (Bachiller 1996, and references
therein). 
In the HH\,7--11 case, the chain of HH objects is more or less coincident
with the axis of the SHV shell, in
agreement with the idea that such Herbig Haro objects are tracing
shocks in the central jet/wind. 

The presence of strong SiO emission is also thought to be a property
of very young outflows.  The strongest component of SiO emission in
the HH\,7--11 vicinity arises from low velocity material, and seems
therefore to be related to ambient gas rather than outflowing material
(Bachiller et al. 1998, Lefloch et al. 1998, Codella et al. 1999). But
as discussed in Section 3.1, there is also EHV SiO emission arising
from at least one of the bullets in the jet. This
SiO emission is quite weak compared to most Class 0 objects, which explains
why it remained undetected in lower resolution observations, but it is
significant.

In summary, although SVS\,13 is a Class~I object (based on its
spectral energy distribution), its outflow presents some features of
outflows from Class~0 sources. Furthermore, it is also remarkable that
the nearby object SVS\,13B, which is at a distance of only 4300 AU
from SVS\,13, is a bona-fide Class~0 object (Bachiller et
al. 1998). At the view of these observational facts, we therefore
suggest that SVS\,13 is at the beginning of the Class~I stage, and
thus keeps some characteristics of the previous evolutionary stage. 
Another possibility is that SVS\,13 is a relatively evolved 
Class\,0 object which has already become 
optically visible due to a favourable viewing angle. We 
note that SVS\,13 is about an order of magnitude more luminous than
typical Class~0 sources, which hints at a more massive object. The
persistence of Class~0 features in the outflow from the early Class~I 
- or ``mature'' Class~0-  source 
SVS\,13 could therefore result from the faster evolution of the
central star.

\begin{figure}[t]
\includegraphics[width=16cm]{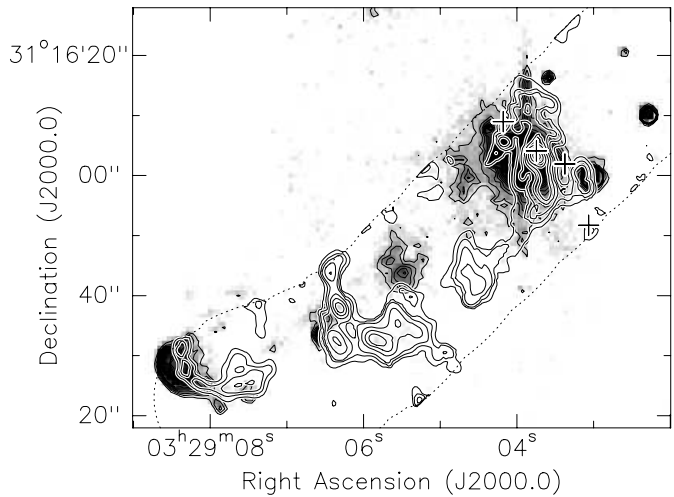}
\caption{Overlay of the K$'$ emission (greyscale; from Hodapp 1994)
and the SHV blueshifted CO $J$=2--1 emission (as in Fig. 2).}
\label{}
\end{figure}

\subsection{The nature of the VLA sources}
The nature of the sources VLA\,3 and VLA\,20 remains uncertain.
Although they are well detected at cm wavelengths (Rodr\'{\i}guez et
al. 1997, 1999), their continuum mm-wave emission -if existent at all-
is very low (Bachiller et al. 1998). The probability of finding
two extragalactic sources so close to SVS\,13 is negligible, and we
should conclude that these objects are associated to the molecular
cloud. The spectral index calculated from the cm to the mm range is
consistent with a flat spectrum, so one possibility could be that they
are ionized knots in the outflows from other sources. However, both
VLA\,3 and VLA\,20 are relatively well aligned with SVS\,13 and
SVS\,13B on a line which is more or less perpendicular to the
direction of the main outflows in the region (those from SVS\,13 and
SVS\,13B), making this interpretation unlikely.

The upperlimit to the $\lambda$\,1.3\,mm flux of VLA\,3 and VLA\,20 is
$\sim$15 mJy which, under the assumptions used by Bachiller et
al. (1998), indicates that a possible dusty circumstellar envelope or
disk around these objects would have less than 0.05~{M$_\odot$} of
mass.  So, if they are young stars, their small amount of
circumstellar mass would in principle indicate that they are
relatively evolved.  Moreover, in such a case, they should have 
very low luminosities, since -in spite of their low circumstellar
extinction- they are not seen in the near-infrared.  More
observations are needed to unveal the nature of these enigmatic VLA
sources. In case they were confirmed to be stellar objects associated
to the molecular cloud, it will be extremely interesting to estimate
their masses and to study their evolutionary status.

\begin{acknowledgements}
We remain obliged to the IRAM staff at Plateau de Bure for help with
the observations, and in particular to those staff members who lost
their lives in two tragic accidents in 1999. RB and MT acknowledge
support from Spanish DGES grant PB96-104.
\end{acknowledgements}


\begin{thebibliography}{}

%   \bibitem[]{} 
%Andr\'e P., Ward-Thompson D., Barsony M., 1993, \ApJ 406, 122
   \bibitem[]{} 
Aspin C., Sandell G., Russel A.P.G., 1994, \AAS 106, 165 
   \bibitem[]{} 
Bachiller R., 1996,  \ARAA 34, 111 
   \bibitem[]{} 
Bachiller R., Cernicharo J., 1990, \AA 239, 276
   \bibitem[]{} 
Bachiller R., Guilloteau S., Gueth F., et al. 1998, \AA 339, L49
%   \bibitem[]{} 
%Bachiller R.,  Mart{\i}n-Pintado J., Fuente A., 1991, \AA 243, L21
   \bibitem[]{} 
Bally J., Devine D., Reipurth B., 1996, \ApJ 473, L49
%   \bibitem[]{} 
%Beckwith et al. 1990
%   \bibitem[]{} 
%Caselli P., Hartquist T.W., Havnes O., 1997, \AA 322, 296
%   \bibitem[]{} 
%Chini R., Reipurth B., Sievers A., et al. 1997, \AA 325, 542
   \bibitem[]{} 
Cernis K., 1990, ApSS 166, 315
   \bibitem[]{} 
Chrysostomou A., Hobson J., Davis C.J., et al. 2000, \MNRAS 314, 229
   \bibitem[]{} 
Codella C., Bachiller R., Reipurth B., 1999, \AA 343, 585
   \bibitem[]{} 
Eisl\"offel J., G\"unther E., Hessman F.V. et al. 1991 \ApJ 383, L19 
%   \bibitem[]{} 
%Gueth F., Guilloteau S., Bachiller R. 1998, \AA 333, 297
%   \bibitem[]{} 
%Grossman E.N., Masson C.R., Sargent A.I., et al. 1987, \ApJ 320, 356
%   \bibitem[]{} 
%Haschick A.D., Moran J.M., Rodr{\i}guez L.F., et al. 1980, \ApJ 237, 
   \bibitem[]{} 
Herbig G.H., Jones B.F., 1983, \AJ 88, 1040
%   \bibitem[]{} 
%Hildebrand 1983
   \bibitem[]{}
Hodapp K.-W., 1994, \ApJS 64, 615
   \bibitem[]{} 
Hodapp K.-W., Ladd E.F., 1995, \ApJ 453, 715
   \bibitem[]{} 
Knee L.B.G., Sandell G., 2000, \AA, in press
%   \bibitem[]{} 
%Lefloch B., Castets A., Cernicharo J., et al. 1998a, \AA 334, 269
   \bibitem[]{} 
Lefloch B., Castets A., Cernicharo J., et al. 1998, \ApJ 504, L109
   \bibitem[]{} 
Lizano S., Heiles C., Rodr{\i}guez, L.F., et al. 1988, \ApJ 328, 763 
   \bibitem[]{} 
Masson C.R., Mundy L.G., Keene J., 1990, \ApJ 357, L25
   \bibitem[]{} 
Mauron N., Thouvenot E., 1991, IAU Circ., 5261
   \bibitem[]{} 
Molinari S., Liseau R., Lorenzetti D., 1993, \AAS 101, 59
   \bibitem[]{} 
Moriarty-Schieven G.H., Snell R.L., Strom S.E., et al. 1987, \ApJ 319, 742
%   \bibitem[]{} 
%Ossenkopf V., Henning T., 1994, \AA 291, 943
%   \bibitem[]{} 
%Pollack J.B., Hollenbach D., Beckwith S., et al. 1994, \ApJ 421, 615
%   \bibitem[]{} 
%Rodr{\i}guez L.F., Cant\'o J. 1983, \RMAA 8, 163
   \bibitem[]{} 
Rodr{\i}guez L.F., Lizano S., Cant\'o J., et al. 1990, \ApJ 365, 261
   \bibitem[]{} 
Rodr{\i}guez L.F., Anglada G., Curiel S., 1997, \ApJ 480, L125
   \bibitem[]{} 
Rodr{\i}guez L.F., Anglada G., Curiel S., 1999, \ApJ SS 125, 427
%   \bibitem[]{} 
%Sandell G., Aspin C., Duncan W.D., et al. 1990, \AA 232, 347
%   \bibitem[]{} 
%Schilke P., Walmsley C.M., Pineau des F\^orets G., et al. 1997, \AA 321, 293  
%   \bibitem[]{} 
%Snell R., Bally J. 1986, ApJ 303, 683
%   \bibitem[]{} 
%Shang H., Shu F.H., Glassgold A.E. 1998, \ApJ 493, L91
   \bibitem[]{} 
Snell R.L., Edwards S. 1981, \ApJ 251, 103
%   \bibitem[]{} 
%Stapelfeldt K.R., Beichman C.A., Hester J.J., et al. 1991, \ApJ 371, 226
   \bibitem[]{} 
Strom, S.E., Vrba, F.J., Strom, K.M. 1976, \AJ 81, 314
%   \bibitem[]{} 
%Woody D.P., Scott S.L., Scoville N.Z., et al. 1989, \ApJ 337, L41

\end{thebibliography}
\end{document}